\documentstyle[prl,aps,psfig]{revtex}

\begin{document}
\draft

\title{Dephasing of Josephson oscillations between two coupled Bose-Einstein condensates}

\author{Pierre Villain$^{1,2}$ and Maciej Lewenstein$^{2}$}

\address{$^1$Commissariat \`a L'Energie Atomique, DSM/DRECAM/SPAM\\  
Centre d'Etudes de Saclay, 91191 Gif-sur-Yvette, France}

\address{$^2$Instit\"ut f\"ur Theoretische Physik, Universit\"at Hannover\\
Appelstr.2, 30167 Hannover, Deutschland}

\maketitle

\begin{abstract}
We study the dynamics of the relative phase between two atomic Bose-Eistein condensates coupled via collisions and via a Josephson-like coupling. We derive the equations of the motion for the relative phase and the relative number operators from the second quantized Hamiltonian of the system using a
quantum field theoretical approach. We distinguish the cases in which the two condensates are in the same trap  or in two different traps and study the influence of this difference on the first order correlation function of atomic fields. In identical traps this function does not undergo dephasing.
We calculate the dephasing time for the case of different traps.  
\end{abstract}


\section{Introduction}

Since the observation  of the Bose-Einstein condensation (BEC) in trapped cold  alkali gases \cite{bec-Rb,bec-Na,bec-Li}, a lot of attention has been devoted to explore the coherence properties of the condensates. 
In particular, in a recent experiment, the interference pattern between the two overlapping condensates has been measured \cite{interf}, in very good agreement with subsequent theoretical calculations \cite{hartmut}.  In the context of interference, the static and  dynamic properties of the phase of the condensate are of  major importance, and for this reason they have been  the object of many theoretical studies in the recent years. First of all, the question  of very existence of the phase has been put forward. The standard 
theory of BEC that is Bogoliubov-Hartree-Fock theory \cite{nozieres} assumes implicitely that the condensation 
in the weakly interacting Bose gas occurs into a {\it coherent} state 
of the lowest energy mode of the system. Such a state admits a defined phase and amplitude, but  exhibits inevitable fluctuations of both these quantities.
In particular, it assumes fluctuations of the number of atoms in the condensate.
On the other hand, if the atom number in a system is conserved, then it is clear that at very low temperature the condensate will be in a {\it Fock}, or in another words {\it number} state, which excludes any atom number fluctuations \cite{gardiner}. Moreover, it is not possible to assign a definite phase to such Fock state. 

Several authors have shown, however, that two overlapping condensates show
similar interference patterns, independently of whether they are in the 
coherent or Fock states  \cite{Javan,Dalib,Cirac}. The reason is that repetitive measurements on two condensates  lead to a build-up of the 
relative atom number fluctuations, and consequently, to a determination of a phase. These results, stimulated by older  discussions on the existence of the phase of the condensate \cite{Leggett}, has led 
to a conclusion that many properties of systems with large number of particles
can be described either way, i.e. using coherent states and their phases, or avoiding it \cite{Molmer}.

The questions what is the character of the quantum state into which the condensation occurs, and what are its properties are nevertheless of major importance. In particular, recently several authors have pointed out that 
the relative phase of two condensates, if established, must necessarily
undergo collapse and revivals due to non-linear dynamics of 
interactions between atoms \cite {Lewen,revue,Walls,jawilk}. The dynamics of the phase reflects the quantum properties of the state of the system, and in particular the collapse time is inversely proportional to the dispersion of the relative number of atoms in the two condensates system.

Originally, the calculation of the phase properties has been done 
 with a single   condensate \cite{Lewen,Walls}. 
In particular, in Ref. \cite{Lewen} we have developed a fully quantum field theoretical method to describe the phase dynamics and to 
 calculate the collapse,  assuming a breaking of the $U(1)$ (phase) 
symmetry of the system. Such a symmetry breaking 
admits the existence of fluctuations of the phase and of the number of atoms in the condensate. The results concerning the dynamics of an absolute phase and absolute atom number for  a single condensate can be straightforwardly applied
to the case of a relative phase and atom number for double, or multiple condensates in stationary states\cite{revue,Atac}.

In fact, in Ref. \cite{revue} we examined two aspects of the 
phase diffusion or collapse in the stationary 
state  of two condensates, that overlap and may even interact through 
elastic collisions. First, we have calculated the properties of the quantum state of such system after a preparation phase, during which the condensates were coupled via coherent Josephson-like coupling. Such coupling leads to fluctuations of the relative atom number, and determines a well
defined relative phase between the condensates in the ground (stationary) state of the system. After the preparation period, the coupling between the condensates is turned off, and the condensates may  evolve freely, or  
remain coupled through the phase and relative number preserving collisions.
In both cases the memory of the prepared phase is lost after
 the collapse time which is now given by a sum of two contributions of the two
 condensates, each inversely proportional to the dispersion of the corresponding atom number.

It should be stressed that although the revival times are typically beyond the reach of present experiments, the collapse phenomenon and its influence 
on the phase memory could be in principle measured. So far, the JILA group 
has realized  two overlapping condensates of rubidium atoms in $(F=1,m_F-1)$ and $(F=2,m_F=2)$ in a magnetic trap \cite{JILA2}. The overlap in such a system is only partial, but nevertheless allows in principle to transfer atoms coherently from one state to another using appropiate sequence of radiofrequency and microwave transitions (alternatively, far resonant stimulated Raman transition could be used). In a more recent paper \cite{cornelphase}, the JILA group reported the results of experiments aiming in testing the robustness of the relative phase between two overlapping condensates. Their results show that the random phase shift due to the coupling with the environment is not significant on a during of the order of 100ms. Conversely the effects of the quantum phase fluctuations were negligible on this time scale. Namely, in the conditions of the JILA experiments ($5.10^5$ atoms in a trap with axial frequency 59 Hz) we estimate the diffusion time to be of the order of 0.5 s. So that the time linear shift due to quantum fluctuations could not be noticeable. It is to be noticed that these effects could be seen  with a condensate formed of less atoms.   
In the recent experiment of the MIT group \cite{Kett2} three co-existing condensates in the ground state of sodium $(F=1; m_F=-1,0,1)$
have been created by transferring the $(F=1,m=-1)$  condensate 
from the magnetic to a purely optical dipole trap. In such trap atoms in 
all Zeeman states remain trapped, and moreover, feel the same trapping potential. Using 
radiofrequency transition one can transfer atoms from one state to another.

Both above mentioned experiments will allow thus to study coherent Josephson 
oscillations between the two, or more condensates. This problem has been 
already studied theoretically, and the corresponding nonlinear Schr\"odinger 
equation for the condensate wave functions has been investigated  numerically \cite{joseph}.
To our knowledge  the problem of quantum fluctuations around dynamical solutions of the coupled Gross-Pitaevskii equations (i.e. the problem of coupled 
time dependent Bogoliubov-de Gennes equations) have not been studied so far.
Some attempts to study the quantum dynamics have only been made for  the 
simplified two mode models \cite{walls}.
In particular, the problem of dephasing of the  Josephson oscillations
 has not been solved  using  a full quantum field theoretical approach. There
are two physical aspects of such dephasing: first, it occurs due the intrinsic
nonlinearity of the Josephson oscillations, that is the fact that oscillation frequency depends, in general, on the inital state of the system. Since any quantum state exhibits  necessarily quantum fluctuations, it is to expect that
the frequency of the corresponding Josephson oscillations would  exhibit 
some fluctuations too, leading thus to dephasing. Second, the quantum nature of the system may lead to a spreading of the distribution of the relative phase
between the two condensates simultaneously  with  the Josephson oscillations.
In this paper we investigate the interplay between those two effects. 

In section 2 we describe our model of two interacting condensates, 
and derive the Heisenberg equations of the atomic field operators. 
Then, using a Bogoliubov approach \cite{nozieres},  we obtain a set of coupled Gross-Pitaevskii equations for the mean values of the fields operators,  and the coupled time-dependent Bogoliubov-de Gennes equations for the quantum  fluctuating parts of the field operators.
In section 3 we introduce the operators describing the fluctuations of the total and relative number of atoms and those describing the total and relative phase, respectively. In section 4 the dynamical equations for these operators are derived using a rotating wave approximation, which allows to decouple the slow evolutions of these operators from the relatively faster evolution of the quasi-particle operators.
In section 5 we solve analytically the dynamics of the phase and atom  number operators in the case when the two condensates are trapped by the same
potential (symmetric case). We show that the first order 
correlation function of the atomic fields of the 
two condensates does not present  any dephasing in time.  
In section 6 we study the case of two different trap potentials (the assymetric case),  and solve the equations 
perturbatively with respect to the relative potential difference. Our study shows that there exists a dephasing, characterized by a time scale whose dependance with the different parameters of the problem is calculated. We conclude in section 7.

\section{Description of the model}
Our system is composed of two condensates consisting of atoms in two different internal states $A$ and $B$ in two  overlapping harmonic traps characterized by the different frequencies $\omega_ {A}$ and $\omega_{B}$. The difference exists in general in a magnetic trap since different Zeeman states react differently to the magnetic field; nevertheless, the frequencies might be the same if the trap is purely optical such as recently used by the MIT group \cite{Kett2}. At very low temperature the interaction between the atoms is well modeled by a contact interaction potential $u_{0} \delta (\vec r -\vec r\ ')$ with $u_{0}= 4 \pi \hbar^{2} a_{SC}/m$ where $a_{SC}$ is the scattering lenght of the collision process. In our case we consider three different scattering processes: the $A-A$, $B-B$ and $A-B$ collisions, and thus three different scattering lenghts. For the case of the Rubidium as in the JILA experiment we  can neglect the differences between these scattering lenghts and assume that
the interaction potential is the same for every atom regardless of its internal state. In the following we will therefore write $u_{AA}=u_{BB}=u_{AB}=u_{0}$. 
We assume also a coherent Josephson like coupling between the A and B atoms which can be realized using a laser (Raman) transition between the two states $A$ and $B$. In order to make this process efficient it is necessary to have a non-negligible overlap between the two condensates. This last condition can be achieved in the JILA case \cite{JILA2} (which will be our reference experiment for the general study of the two differents traps), as well as in the optical trap experiment of MIT .

The second quantized Hamiltonian of the system is:
\begin{eqnarray}
{\cal H}&=& \int\! d\vec r\,\hat\Psi_A^{\dag}(\vec r\,,t)
\left[-\frac{\hbar^2}{2M}\nabla^2+V_A(\vec r\,)\right]\hat\Psi_A(\vec r\,,t) \nonumber\\
&+&\frac{u_{0}}{2} \int\! d\vec r\, 
\hat\Psi_A^{\dag}(\vec r\,,t)\hat\Psi_A^{\dag}(\vec r\,,t)
\hat\Psi_A(\vec r\,,t)\hat\Psi_A(\vec r\,,t) \nonumber\\
&+&\int\! d\vec r\,\hat\Psi_B^{\dag}(\vec r\,,t)
\left[-\frac{\hbar^2}{2M}\nabla^2+V_B(\vec r\,)\right]\hat\Psi_B(\vec r\,,t) \nonumber\\
&+&\frac{u_{0}}{2}\int\! d\vec r\, 
\hat\Psi_B^{\dag}(\vec r\,,t)\hat\Psi_B^{\dag}(\vec r\,,t)
\hat\Psi_B(\vec r\,,t)\hat\Psi_B(\vec r\,,t)\nonumber\\
&+&u_{0}\int\! d\vec r\, 
\hat\Psi_A^{\dag}(\vec r\,,t)\hat\Psi_B^{\dag}(\vec r\,,t)
\hat\Psi_B(\vec r\,,t)\hat\Psi_A(\vec r\,,t)\nonumber \\
&-&\hbar \lambda \int\! d\vec r\,\left[\hat\Psi_B^{\dag}(\vec r\,,t)\hat\Psi_A(\vec r\,,t)
+ \hat\Psi_A^{\dag}(\vec r\,,t)\hat\Psi_B(\vec r\,,t)\right]
\label {dupa} ,
\end {eqnarray}
where $\hat\Psi_{A,B}(\vec r\,,t)$ denote the atomic field operators fullfilling the standard (bosonic) commutation relations:
\begin{eqnarray}
\left[ \hat\Psi_{A,B}(\vec r\,,t),\hat\Psi^{\dag}_{A,B}(\vec r^{\ '},t) \right]=\delta (\vec r-\vec r^{\ '} ).
\end{eqnarray}
The last term in the Hamiltonian (\ref{dupa}) represents the Josephson like coupling which allows exchange of atoms from one state to another. The frequency $\lambda$ characterizes the strenght of the coupling between the two atomic states and its value can be experimentally controlled.

From the Hamiltonian (\ref{dupa}) we derive the Heisenberg equations for the atomic field operators:
\begin{eqnarray}
i\hbar \frac{\partial \hat\Psi_A(\vec r\,,t)}{\partial t} &=&[\hat\Psi_A(\vec r\,,t),\cal H] \nonumber \\
&=& \left[-\frac{\hbar^2}{2M}\nabla^2+V_A(\vec r\,)\right]\hat\Psi_A +u_{0}\hat\Psi^{\dag}_A\hat\Psi_A\hat\Psi_A
+u_{0}\hat\Psi^{\dag}_B\hat\Psi_B\hat\Psi_A
-\hbar \lambda u_{0}\hat\Psi_B ,
\label {HeisenA}
\end{eqnarray}
and 
\begin{eqnarray}
i\hbar \frac{\partial \hat\Psi_B(\vec r\,,t)}{\partial t} &=&[\hat\Psi_B(\vec r\,,t),\cal H] \nonumber \\
&=& \left[-\frac{\hbar^2}{2M}\nabla^2+V_B(\vec r\,)\right]\hat\Psi_B +u_{0}\hat\Psi^{\dag}_B\hat\Psi_B\hat\Psi_B
+u_{0}\hat\Psi^{\dag}_A\hat\Psi_A\hat\Psi_B
-\hbar \lambda u_{0}\hat\Psi_A .
\label{HeisenB}
\end{eqnarray}
We use then the well-known time dependent Bogoliubov method to find the dynamics of the system at $T=0$. To this aim we assume that the atomic fields operators can be written as :
\begin{eqnarray}
\hat\Psi_A(\vec r\,,t)= \sqrt{N}\psi_{A}(\vec r\,,t) +\delta \hat \psi_{A}(\vec r\,,t) , \\
\hat\Psi_B(\vec r\,,t)= \sqrt{N}\psi_{B}(\vec r\,,t) +\delta \hat \psi_{B}(\vec r\,,t) ,
\end{eqnarray}
where $N$ denotes the total number of atoms, $\sqrt{N}\psi_{A}(\vec r\,,t)$ and $\sqrt{N}\psi_{B}(\vec r\,,t)$ are in fact the wave functions of the $A$ and $B$ condensates whereas the $\delta \hat\psi_{A,B}$ denote the fluctuating parts of the operators, respectively.
By setting these expressions into the Eqns (\ref{HeisenA}) and (\ref{HeisenB}) and identifying the parts of zero and first order respect to the fluctuations, we find the following set of coupled equations:
\begin{mathletters}
\begin{eqnarray}
i\hbar \frac{\partial \psi_{A}(\vec r\,,t)}{\partial t} &=& {\cal L }_{A} \psi_{A}(\vec r\,,t)+ u_{0} (\rho_{A} +\rho_{B})\psi_{A}(\vec r\,,t)-\hbar \lambda \psi_{B}(\vec r\,,t) , \label{NLSE1} \\
i\hbar \frac{\partial \psi_{B}(\vec r\,,t)}{\partial t}&=& {\cal L}_{B} \psi_{B}(\vec r\,,t)+ u_{0} (\rho_{A} +\rho_{B})\psi_{B}(\vec r\,,t)-\hbar \lambda \psi_{A}(\vec r\,,t) , \label{NLSE2}
\end{eqnarray}
\end{mathletters}
for the zero order part. Here ${\cal L}_{A,B} = \left[-\frac{\hbar^2}{2M}\nabla^2+V_{A,B}(\vec r\,)\right]$, while $\rho_{A,B}= N|\psi_{A,B}(\vec r\,,t)|^2$ are the densities of the $A$ and $B$ atoms. In the following we will denote by $\rho =\rho_{A} +\rho_{B}$ the total density of atoms. Moreover we assume the following normalisation conditions:$\int\! d^3\vec r \  N|\psi_{A,B}(\vec r\,,t)|^2= N_{A,B}(t)$ where $N_A$ and $N_B$ denote the number of the A and B atoms. Obviously in this case we have $\int\! d^3\vec r \  [|\psi_{A}(\vec r\,,t)|^2+|\psi_{B}(\vec r\,,t)|^2]=1$ so that the total number of atoms is fixed to $N$. Eqs. (\ref{NLSE1},\ref{NLSE2}) constitute a system of coupled Gross-Pitaevskii equations.

In the first order in the fluctuating parts we find:
\begin{mathletters}
\begin{eqnarray}
i \hbar \frac{\partial \delta \hat \psi_{A}(\vec r\,,t)}{\partial t} &=& \left[{\cal L }_{A}+u_{0}\rho \right]\delta \hat \psi_{A} + Nu_{0}\psi_{A}\left[\psi_{A}\delta \hat \psi^{\dag}_{A} + \psi^{*}_{A}\delta \hat \psi_{A}\right]+Nu_{0}\psi_{A}\left[\psi_{B}\delta \hat \psi^{\dag}_{B} + \psi^{*}_{B}\delta \hat \psi_{B}\right] 
- \hbar \lambda \delta \hat \psi_{B} \label{bog1}, \\
i \hbar \frac{\partial \delta \hat \psi_{B}(\vec r\,,t)}{\partial t} &=& \left[{\cal L }_{B}+u_{0}\rho \right]\delta \hat \psi_{B} + Nu_{0}\psi_{B}\left[\psi_{B}\delta \hat \psi^{\dag}_{B} + \psi^{*}_{B}\delta \hat \psi_{B}\right] 
+Nu_{0}\psi_{B}\left[\psi_{A}\delta \hat \psi^{\dag}_{A} + \psi^{*}_{A}\delta \hat \psi_{A}\right] 
- \hbar \lambda \delta \hat \psi_{A} \label{bog2} .
\end{eqnarray}
\end{mathletters}
These equations in turn constitute a system of coupled Bogoliubov-de Gennes equations (\cite{nozieres,de Gennes}).
As the Hamiltonian is invariant under the permutation of the $A$ and $B$ atoms, it is convenient to introduce the following atomic field operators:
\begin{mathletters}
\begin{eqnarray}
\hat \Psi_+ &=& e^{-i\lambda t} [\hat \Psi_A+\hat \Psi_B]/ \sqrt 2 \label{new1},  \\
\hat \Psi_- &=& e^{i\lambda t}  [\hat \Psi_A-\hat \Psi_B]/ \sqrt 2 \label{new2}.
\end{eqnarray}
\end{mathletters}
Using the  previous evolution equations we can easily derive the equations for the new wave functions $\psi_+$ and $\psi_-$:

\begin{mathletters}
\begin{eqnarray}
i\hbar \frac{\partial \psi_+(\vec r\,,t)}{\partial t} &=& \left[{\cal L}+u_0\rho\right]\psi_+ +\delta\tilde V \psi_-  , \label{NNLSE1}\\
i\hbar \frac{\partial \psi_-(\vec r\,,t)}{\partial t} &=& \left[{\cal L}+u_0\rho\right]\psi_- +\delta\tilde V^* \psi_+ ,\label{NNLSE2}
\end{eqnarray}
\end{mathletters}
where ${\cal L}=\left[-\frac{\hbar^2}{2M}\nabla^2+V_m(\vec r\,)\right]$, $\delta\tilde V = \delta V e^{-2i\lambda t}$,  $V_m=(V_A + V_B)/2$, $\delta V =(V_A - V_B)/2$. Notice that the unitary character of the transformation (\ref{new1},\ref{new2}) leaves the total density of atoms invariant: $\rho =\rho_{A} +\rho_{B}=\rho_+ +\rho_-$, with $\rho_{\pm} = N|\psi_{\pm}|^2$.
For the fluctuating parts of the operators we obtain :

\begin{mathletters}
\begin{eqnarray}
i\hbar \frac{\partial \delta \hat \psi_+(\vec r\,,t)}{\partial t} &=& \left[{\cal L }+u_{0}(2\rho_+ +\rho_-) \right]\delta \hat \psi_+ + \delta \tilde V \delta \hat \psi_- +Nu_0 \psi_+^2 \delta \hat\psi^{\dag}_+ + Nu_0 \psi_+ ( \psi_- \delta\hat \psi^{\dag}_- + \psi^*_- \delta\hat \psi_-)  , \label{ndelpsi1} \\
i\hbar \frac{\partial \delta \hat \psi_-(\vec r\,,t)}{\partial t} &=& \left[{\cal L }+u_{0}(\rho_+ +2\rho_-) \right]\delta \hat \psi_- + \delta \tilde V^* \delta \hat \psi_+ +Nu_{0}\psi_-^2 \delta \hat\psi^{\dag}_- + Nu_0 \psi_- ( \psi_+ \delta\hat \psi^{\dag}_+ + \psi^*_+ \delta\hat \psi_+). \label{ndelpsi2}
\end{eqnarray}
\end{mathletters}

\section{Total atom number and phase}
In this section we introduce the operators describing the total atom number and phase fluctuations. We are following the method previously developed in refs.\cite{Lewen,revue} to study the same system, but in the static case.
We begin by introducing the following hermitian operators:
\begin{mathletters}
\begin{eqnarray}
\hat P_+ &=& \int (\psi^*_+ \delta \hat \psi_+ +\psi_+ \delta \hat \psi^{\dag}_+) d^3 r , \label{p+}\\
\hat P_- &=& \int (\psi^*_- \delta \hat \psi_- +\psi_- \delta \hat \psi^{\dag}_-) d^3 r \label{p-} .
\end{eqnarray}
\end{mathletters}
These operators describe, respectively, the fluctuations of the number of atoms in the new states $\psi_+$ and $\psi_-$.
Using the equations (\ref{NNLSE1},\ref{NNLSE2},\ref{ndelpsi1},\ref{ndelpsi2}) we obtain the evolution equations of this two operators in a straighforward way:
\begin{mathletters}
\begin{eqnarray}
i\hbar \frac{d \hat P_+}{dt} &=& \int \delta \tilde V\left[\psi_- \delta \hat \psi^{\dag}_+ + \psi^*_+ \delta \hat \psi_- \right] d^3 r - \int \delta \tilde V^* \left[\psi^*_- \delta \hat \psi_+ + \psi_+ \delta \hat \psi^{\dag}_- \right] d^3 r , \label{evoluP1}\\
i\hbar \frac{d \hat P_-}{dt} &=& -i\hbar \frac{d \hat P_+}{dt} .
\label{evoluP2} 
\end{eqnarray}
\end{mathletters}
We then construct the operators:
\begin{mathletters}
\begin{eqnarray}
\hat Q_+ &=& i\hbar \int (\phi^*_+ \delta \hat \psi_+ -\phi_+ \delta \hat \psi^{\dag}_+) d^3 r ,\label{Q1}\\
\hat Q_- &=& i\hbar \int (\phi^*_- \delta \hat \psi_- -\phi_- \delta \hat \psi^{\dag}_-) d^3 r , 
\label{Q2}
\end{eqnarray}
\end{mathletters}
for which the commutation relations read:
\begin{eqnarray}
\left[\hat Q_+ , \hat P_+ \right] &=& 2i\hbar{\rm Re}J_+ , \\
\left[\hat Q_- , \hat P_- \right] &=& 2i\hbar{\rm Re} J_- , 
\end{eqnarray}
where
\begin{mathletters}
\begin{eqnarray}
J_+ &=& \int \phi^*_+ \psi_+ d^3 r , \\
J_- &=& \int \phi^*_- \psi_- d^3 r .
\end{eqnarray}
\end{mathletters}
In the following we will denote ${\rm Re}J_{\pm}$ by $\gamma_{\pm}$.

The operators (\ref{Q1},\ref{Q2}) up to a normalisation factor are conjugated to the operators (\ref{p+},\ref{p-}) and describe the fluctuations of the phases of the new atomic field operators. In order to study their dynamics, we have to find the evolution equations for the functions $\phi_+$ and $\phi_-$. For this purpose we introduce the operators describing the fluctuations of the total number of atoms and the total phase of the system:
\begin{eqnarray}
\hat P_{tot} &=& (\hat P_+ + \hat P_-)/\sqrt 2 , \\
\hat Q_{tot} &=& (\hat Q_+ + \hat Q_-)/\sqrt 2 ,
\end{eqnarray}
The latter operator fulfills ( \cite{Lewen,revue}):
\begin{eqnarray}
\left[\hat Q_{tot} , \hat P_{tot} \right] = i\hbar,\label{normatot}
\end{eqnarray}
and
\begin{eqnarray}
\frac{d\hat Q_{tot}}{dt} = N\tilde u \hat P_{tot},\label{tot}
\end{eqnarray}
where the coefficient $\tilde u$ will be define in the following.
The commutation relation (\ref{normatot}) imposes the following relation:
\begin{eqnarray}
{\rm Re}J_+ + {\rm Re}J_- = 1\label{norma1}
\end{eqnarray} 
and the operator evolution equation (\ref{tot}) gives the evolution equations for $\phi_+$ and $\phi_-$:
\begin{mathletters}
\begin{eqnarray}
i\hbar \frac{\partial \phi_+(\vec r\,,t)}{\partial t} &=& \left[{\cal L}+u_0(2\rho_+ + \rho_-)\right]\phi_+ +Nu_0\psi^2_+ \phi^*_+ + Nu_0 \psi_+ (\psi_- \phi^*_- + \psi^*_- \phi_-) - N\tilde u \psi_+ + \delta\tilde V \phi_-  ,\label{evoluphi1}\\
i\hbar \frac{\partial \phi_-(\vec r\,,t)}{\partial t} &=& \left[{\cal L}+u_0(\rho_+ + 2\rho_-)\right]\phi_- +Nu_0\psi^2_- \phi^*_- + Nu_0 \psi_- (\psi_+ \phi^*_+ + \psi^*_+ \phi_+) - N\tilde u \psi_- + \delta\tilde V^* \phi_+ .
\label{evoluphi2}
\end{eqnarray}
\end{mathletters}
In order to define in a unique way the functions $\phi_+$ and $\phi_-$ and the coefficient $\tilde u$ (indeed the functions  $\phi_{\pm}' = \phi_{\pm}+i\alpha \psi_{\pm}$ where $\alpha$ is a real coefficient are solutions of the above equations too)  we impose a more restrictive normalisation condition by requiring additionally:
\begin{eqnarray}
\int (\phi^*_+ \psi_+ + \phi_- \psi^*_-) d^3 r &=& 1 ,
\end{eqnarray}
{\it id est}
\begin{eqnarray}
J_+ + J_- = 1.\label{norma2}
\end{eqnarray}

It is interesting to note that there exists a direct relation between $\phi_{\pm}$ and $\psi_{\pm}$, namely:
\begin{eqnarray}
\phi_{\pm} &=& 2N[\frac{\partial \psi_{\pm}}{\partial N}-\frac{i}{\hbar}f(t) \psi_{\pm}] ,
\end{eqnarray} 
with 
\begin{eqnarray}
f (t) &=& \frac{\hbar}{2iN}+\hbar {\rm Im}\int[ \psi^*_+\frac{\partial \psi_+}{\partial N}+ \psi^*_-\frac{\partial \psi_-}{\partial N}] d^3 r ,
\end{eqnarray}
and 
\begin{eqnarray}
\tilde u = -2\frac{df(t)}{dt}= -2\hbar\frac{d}{dt}\left[{\rm Im} \int [ \psi^*_+\frac{\partial \psi_+}{\partial N}+ \psi^*_-\frac{\partial \psi_-}{\partial N}] d^3 r \right] .
\end{eqnarray}

The exact Heisenberg equations for the phase operators have the form :
\begin{mathletters}
\begin{eqnarray}
\frac{d\hat Q_+}{dt} &=& N\tilde u \hat P_+ + Nu_0 \int (\psi_+ \phi^*_+ + \psi^*_+ \phi_+)(\psi_- \delta \hat \psi^{\dag}_- + \psi^*_- \delta \hat \psi_-) d^3 r - Nu_0\int (\psi_- \phi^*_- + \psi^*_- \phi_-)(\psi_+ \delta \hat \psi^{\dag}_+ + \psi^*_+ \delta \hat \psi_+) d^3 r \nonumber \\
&+& \int \delta \tilde V (\phi^*_+ \delta \hat \psi_- - \phi_- \delta \hat \psi^{\dag}_+) d^3 r + \int \delta \tilde V^* (\phi_+ \delta \hat \psi^{\dag}_- -\phi^*_- \delta \hat \psi_+) d^3 r  , \label{evoluQ1}\\
\frac{d\hat Q_-}{dt} &=&  N\tilde u \hat P_- - Nu_0 \int (\psi_+ \phi^*_+ + \psi^*_+ \phi_+)(\psi_- \delta \hat \psi^{\dag}_- + \psi^*_- \delta \hat \psi_-) d^3 r + Nu_0\int (\psi_- \phi^*_- + \psi^*_- \phi_-)(\psi_+ \delta \hat \psi^{\dag}_+ + \psi^*_+ \delta \hat \psi_+) d^3 r \nonumber \\
&-& \int \delta \tilde V (\phi^*_+ \delta \hat \psi_- - \phi_- \delta \hat \psi^{\dag}_+) d^3 r - \int \delta \tilde V^* (\phi_+ \delta \hat \psi^{\dag}_- -\phi^*_- \delta \hat \psi_+) d^3 r .\label{evoluQ2}
\end{eqnarray}
\end{mathletters}

\section{Evolution of the operators in the Rotating Wave Approximation}
It is possible to simplify the previous evolution equations by making a Rotating Wave Approximation (RWA). Namely, we make use the fact that the quasi-particles operators describing the elementary excitations evolve with a characteristic  time of the order of magnitude of the period of the harmonic trap. On the other hand the evolution of the operators $\hat P_+,\hat P_- ,\hat Q_+,\hat Q_-$ is slower, and it is thus possible to project the evolution equations (\ref{evoluP1},\ref{evoluP2}and \ref{evoluQ1},\ref{evoluQ2}) on $\hat P_+,\hat P_- ,\hat Q_+,\hat Q_-$ .
Introducing the following notation:
\begin{eqnarray}
(\phi |\psi) = \int \delta \tilde V \phi^* \psi \ d^3 r ,
\end{eqnarray}
we then find in the RWA :
\begin{mathletters}
\begin{eqnarray}
\frac{d \hat P_+}{dt} &=& \frac{1}{\hbar\gamma_+} {\rm Im}(\phi_+ |\psi_-)\hat P_+ + \frac{1}{\hbar\gamma_-} {\rm Im}(\psi_+ |\phi_-)\hat P_- + \frac{1}{\hbar^2} {\rm Re}(\psi_+ |\psi_-)\left(\frac{\hat Q_+}{\gamma_+} - \frac{\hat Q_-}{\gamma_-}\right) ,\label{P+}\\
\frac{d \hat P_-}{dt} &=& -\frac{d \hat P_+}{dt} ,\label{P-}\\
\frac{d \hat Q_+}{dt} &=& \left[N\tilde u -\frac{2Nu_0 {\cal I}}{\gamma_+} - \frac{1}{\gamma_+} {\rm Re} (\phi_+ |\phi_-)\right]\hat P_+ +\frac{1}{\gamma_-}\left[2Nu_0 {\cal I} + {\rm Re}(\phi_+ |\phi_-)\right]\hat P_-  \nonumber \\
&+&\frac{1}{\hbar\gamma_+} {\rm Im}(\psi_+ |\phi_-)\hat Q_+ + \frac{1}{\hbar\gamma_-}{\rm Im}(\phi_+ |\psi_-)\hat Q_- ,\label{Q+}\\
\frac{d \hat Q_-}{dt} &=& \frac{1}{\gamma_+}\left[2Nu_0{\cal I} + {\rm Re}(\phi_+ |\phi_-)\right]\hat P_+ +\left[N\tilde u -\frac{2Nu_0 {\cal I}}{\gamma_-} - \frac{1}{\gamma_-} {\rm Re} (\phi_+ |\phi_-)\right]\hat P_-  \nonumber \\
&-&\frac{1}{\hbar\gamma_+} {\rm Im}(\psi_+ |\phi_-)\hat Q_+ - \frac{1}{\hbar\gamma_-}{\rm Im}(\phi_+ |\psi_-)\hat Q_-, \label{Q-} 
\end{eqnarray}
\end{mathletters}
where ${\cal I} = \int {\rm Re}(\psi_- \phi^*_-){\rm Re}(\psi_+ \phi^*_+) d^3 r$.
At this stage it is interesting to introduce the operators describing the fluctuations of the relative number of atoms between the states $\psi_+$ and $\psi_-$ and of the relative phase between these two states. We use the approach previously developped in \cite{revue} to define these operators:
\begin{eqnarray}
\hat P_{rel} &=& (\gamma_-\hat P_+ - \gamma_+\hat P_-)/\sqrt 2, \\
\hat Q_{rel} &=& \left(\frac{\hat Q_+}{\gamma_+} - \frac{\hat Q_-}{\gamma_-}\right)/\sqrt 2.
\end{eqnarray}
We can then write the evolution equations of the new set of operators $\hat P_{tot},\hat P_{rel},\hat Q_{tot},\hat Q_{rel}$ in terms of those operators:
\begin{mathletters}
\begin{eqnarray}
\frac{d \hat P_{tot}}{dt} &=& 0 ,\label{Evolutot1}\\
\frac{d \hat Q_{tot}}{dt} &=& N \tilde u \hat P_{tot}, 
\label{Evolutot2}\\
\frac{d \hat P_{rel}}{dt} &=&  \frac{1}{\hbar}\left[\frac{1}{\gamma_+}{\rm Im}(\phi_+ |\psi_-)-\frac{1}{\gamma_-} {\rm Im}(\psi_+ |\phi_-)\right]\hat P_{rel} + +\frac{1}{\hbar^2}{\rm Re}(\psi_+ |\psi_-)\hat Q_{rel} , \label{Evolurel1}\\
\frac{d \hat Q_{rel}}{dt} &=&\frac{1}{\gamma_+ \gamma_-} \left[N \tilde u -\frac{2Nu_0 {\cal I}}{\gamma_+\gamma_-} -\frac{1}{\gamma_+\gamma_-} {\rm Re}(\phi_+ |\phi_-)\right] \hat P_{rel}+ \frac{1}{\hbar}\left[ \frac{1}{\gamma_-}{\rm Im}(\psi_+ |\phi_- ) - \frac{1}{\gamma_+}{\rm Im}(\phi_+ |\psi_- )\right] \hat Q_{rel} \label{Evolurel2}.
\end{eqnarray}
\end{mathletters}
These equations constitute one of the basic results of the present paper.We can remark the  complete decoupling between the evolution of the observables of the total dynamics and these of the relative dynamics.
We shall now discuss the solutions in various limiting cases.

\section{The symmetric case}
We call the symetric case the case where the two condensates $A$ and $B$ are in the same trap (mathematically that means that $\delta V =0$). Experimentaly such situation would occur when the atoms are confined in a purely optical trap. This kind of trap for Bose condensates has been recently realized by the MIT group \cite{Kett2}, and is very important from the theoritical point of view since the problem in this case can be exactly solved. In the, so-called, Thomas-Fermi limit  one can show that the solutions of the Gross-Pitaevskii equations have the form \cite{revue,StringariTF,Schlyap}:
\begin{mathletters}
\begin{eqnarray}
\psi^{(0)}_+ (\vec r,t) &=& \frac{\sqrt{N_+}}{N}\sqrt{\rho (\vec r,t)}e^{i\theta_+(\vec r,t)},\label{zero+} \\
\psi^{(0)}_- (\vec r,t) &=& \frac{\sqrt{N_-}}{N}\sqrt{\rho (\vec r,t)}e^{i\theta_-(\vec r,t)},\label{zero-}
\end{eqnarray}
\end{mathletters}
where the density $\rho$ is the Thomas-Fermi solution of the single condensate Gross-Pitaevskii equation : 
\begin{eqnarray}
\rho =  \frac{15N}{8\pi r_0^3(t)}\left(1-\frac{r^2}{r^2_0(t)}\right),
\end{eqnarray}
and $r_0$ is the size of both condensates in the Thomas-Fermi approximation, fullfilling the equation:
\begin{eqnarray}
\ddot r_0+\omega_t^2(t)r_0 - \frac{15Nu_0}{4\pi mr_0^4}=0.
\end{eqnarray}
The phase $\theta_{\pm}$ has the form $(A(t)r^2 +B_{\pm}(t))/\hbar$, where the coefficients $A(t)$ and $B_{+,-}(t)$ follow the evolution equations previously found in Ref.\cite{revue}:
\begin{eqnarray}
 A &=& -m\dot r_0/2r_0, \\
\dot B_{\pm}&=& 15Nu_0/8\pi r_0^3, 
\end{eqnarray}
 and $N_+$ and  $N_-$ are constant.
We find then the following form for  the functions $\phi_+$ and $\phi_-$:
\begin{eqnarray}
\phi^{(0)}_+ &=& \frac{\alpha (t)}{\psi^{(0)*}_+} , \\
\phi^{(0)}_- &=& \frac{\alpha (t)}{\psi^{(0)*}_-} ,
\end{eqnarray}
where the function $\alpha (t)$ is equal to $(\frac{8 \pi}{3}r^3_0(t))^{-1}$ because of the normalisation conditions (\ref{norma1},\ref{norma2}), which requires more $\gamma^{(0)}_+ =\gamma^{(0)}_- = \frac{1}{2}$.
 Moreover, we have $4u_0 \alpha (t) = \tilde u^{(0)}$ because the functions $\phi^{(0)}_{\pm}$ are solutions of the same equations as $\psi^{(0)}_{\pm}$(see \cite{revue}).
Taking $\delta V =0$ in the equations (\ref{Evolutot1},\ref{Evolutot2},\ref{Evolurel1},\ref{Evolurel2}) and replacing the solutions obtained of the hydrodynamic limit in $\cal {I}$ we get immediatly:
\begin{eqnarray}
\frac{d \hat P_{tot}}{dt} &=& 0 ,\\
\frac{d \hat Q_{tot}}{dt} &=& N \tilde u^{(0)} \hat P_{tot} \label{Qtot},\\
\frac{d \hat P_{rel}}{dt} &=& 0 ,\\
\frac{d \hat Q_{rel}}{dt} &=& 0 .\\
\end{eqnarray}
The fact that $\hat P_{tot}$ and $\hat P_{rel}$ are constants of the movement is related to the conservation of the total number of atoms, and to the conservation of the relative number of atoms (which is evident since both $N_+$ and  $N_-$ are constant).
Moreover, we conclude that the relative phase between the two states $+$ and $-$ is fixed (up to fluctuations constant in time) and does not diffuse. Only the total phase operator $\hat Q_{tot}$ exhibits a linear growth in time (see Eq. (\ref{Qtot})) which leads to a gaussian decay of the correlation function of the type $<\hat\Psi^{\dag}_A(\vec r,t+\tau)\hat\Psi_A(\vec r,t)>$ as discussed in \cite{Lewen}. Using the inverse transformation :
\begin{mathletters}
\begin{eqnarray} 
\hat \Psi_A &=& [e^{i\lambda t}\hat \Psi_+ + e^{-i\lambda t}\hat \Psi_-]/\sqrt 2 ,   \\
\hat \Psi_B &=& [e^{i\lambda t}\hat \Psi_+ - e^{-i\lambda t}\hat \Psi_-]/\sqrt 2  ,
\end{eqnarray}
\end{mathletters}
it is then possible to calculate the one time correlation function $<\hat \Psi^{\dag}_A(\vec r,t) \hat \Psi_B(\vec r,t)>$. This function oscillates with the frequency $2 \lambda $, which means that the fluctuations of the relative phase between these two states don't show any effect of phase diffusion. This is a very interesting  result since it immediatly implies that in weakly asymetric case, the phase diffusion rate will have to scale with $\delta V$.

It is worth noticing that in this particularly simple case it is not necesary to make the transformation (\ref{new1},\ref{new2}). The Gross-Pitaevskii equations for $\psi_A$ and $\psi_B$ can be exactly solved in the hydrodynamic limit using a selfsimilar solution of the form :
\begin{eqnarray}
\psi_A(\vec r,t) &=& \left(\frac{15N_A (t)}{8\pi Nr_0^3(t)}\right)^{\frac{1}{2}}\ \left( 1-\frac{r^2}{r^2_0(t)}\right)^{\frac{1}{2}}\ e^{i(A(t)r^2 +B_A(t))/\hbar} \nonumber \\
\Psi_B(\vec r,t) &=& \left(\frac{15N_B(t)}{8\pi Nr_0^3(t)}\right)^{\frac{1}{2}}\ \left( 1-\frac{r^2}{r^2_0(t)}\right)^{\frac{1}{2}}\ e^{i(A(t)r^2 +B_B(t))/\hbar}
\end{eqnarray}
Introducing the notations $\Delta \Phi$ for the relative phase $\Phi_A - \Phi_B=(B_A(t)-B_B(t))/\hbar$ and $\Delta N$ for the difference of populations $N_A(t)-N_B(t)$, and using the equations of the motion we find:
\begin{mathletters}
\begin{eqnarray}
\Delta \dot N &=& \frac{\partial{\cal C}}{\partial \Delta \Phi}, \label{symeq1}\\
\Delta \dot \Phi &=& -\frac{\partial{\cal C}}{\partial \Delta N},
\label{symeq2}
\end{eqnarray}
\end{mathletters}
where $\cal C$ is given by:
\begin{eqnarray}
{\cal C} = 2\hbar\lambda \sqrt{N^2 -\Delta N^2}\cos(\Delta \Phi).
\end{eqnarray}
The Eqs. (\ref{symeq1},\ref{symeq2}) constitute a set of classical Hamilton equations  for which the function $\cal C$ is a Hamiltonian, i.e. a constant of the motion. By straightforward integration we obtain:
\begin{eqnarray}
\Delta N = {\cal{A}} \cos(2\lambda t +\Phi_{N0}),
\end{eqnarray}
where 
\begin{eqnarray}
{\cal{A}} &=& \sqrt{N^2 -\frac{{\cal{C}}^2}{4\hbar^2 \lambda^2}}, \nonumber \\
  &=& \sqrt{N^2 \sin^2 (\Delta \Phi) +\Delta N^2 \cos^2 (\Delta \Phi)}.
\end{eqnarray}
and $\Phi_{N0}$ is a constant depending of the initial conditions, namely we have $\Phi_{N0}= \arccos(\Delta N(0)/{\cal A})$.
The difference of populations between the two condensates oscillates at the previsible frequency $2\lambda$.
In the same way we find that
\begin{eqnarray}
\tan(\Delta \Phi - \Delta \Phi (0)) = \frac{4\hbar\lambda \cal{A}}{\cal C}\sin(2\lambda t). 
\end{eqnarray}

\section{The asymmetric case}
We want now evaluate the influence of the difference between the two potential traps on the dynamics of the relative phase of the condensates. To this aim we focus our attention on the evolution of the operator $\hat Q_{rel}$, since for the first order correlation function in the Rotating Wave Approximation:
\begin{eqnarray}
<\hat\Psi^{\dag}_+(t)\hat\Psi_-(t)> &\propto & <{\rm exp}\left[i\frac{\sqrt{2}\hat Q_{rel}(t)}{\hbar}\right]> \nonumber \\
				&\propto & {\rm exp}\left[-\frac{1}{\hbar^2}<\hat Q^2_{rel}(t)>\right]
\end{eqnarray}
for a Gaussian quantum stochastic process (see \cite{qn}).
Knowing that the fluctuations of the numbers of the atoms induce phase diffusion (see \cite{Lewen},\cite{revue}), we can study the influence of the respective fluctuations of the total number of atoms and of the relative number of atoms on the evolution of the relative phase.
In fact the evolution of $\hat Q_{rel}$ doesn't depend of $\hat P_{tot}$ as it can be seen in the Eqns. (\ref{Evolutot1}, \ref{Evolutot2}, \ref{Evolurel1}, \ref{Evolurel2}) so that the fluctuations of the total number of atoms can induce diffusion only for the total phase. We will show that the fluctuations of the relative number of atoms will, in counterpart, induce a diffusion of the relative phase.
We consider the case when $\delta V$ is not equal to zero, but small enough to make a pertubative development of all the interesting physical quantities. We assume that the traps have different frequencies so that $\delta V = m(\omega_A^2 - \omega_B^2 )r^2 /4= m\delta \omega^2 r^2 /4$. In fact the real dimensionless parameter of the perturbative development appears naturally in the calculation and is $\frac{m \delta \omega ^2 r^2_0 /2}{2\hbar \lambda}$ that we will denote $v$ in the following and which represents the ratio between the two characteristic energies of the system.
We solve the equations (\ref{Evolurel1},\ref{Evolurel2}) in the first order in $v$.
To find the first order correction for the relative phase operator, $\hat Q^{(1)}_{rel}$, we have then to calculate some of the integrals appearing in in its evolution equation with the wave functions given at zero order  in the hydrodynamic limit (namely $\psi_{0+},\psi_{0-},\phi_{0+},\phi_{0-}$), some others using the first order correction of these functions (in the hydrodynamic limit) and replace the operators by their contribution of zero order in $\delta V$.\\
In this manner we obtain:
\begin{eqnarray}
\frac{d \hat Q^{(1)}_{rel}}{dt} &=&\frac{1}{\gamma^{(0)}_+ \gamma^{(0)}_-} \left[N \tilde u^{(1)} -\frac{2Nu_0 {\cal I}^{(1)}}{\gamma^{(0)}_+\gamma^{(0)}_-} -\frac{1}{\gamma^{(0)}_+\gamma^{(0)}_-} {\rm Re}(\phi^{(0)}_+ |\phi^{(0)}_-)\right] \hat P^{(0)}_{rel} \nonumber \\
&+& \frac{1}{\hbar}\left[ \frac{1}{\gamma^{(0)}_-}{\rm Im}(\psi^{(0)}_+ |\phi^{(0)}_- ) - \frac{1}{\gamma^{(0)}_+}{\rm Im}(\phi^{(0)}_+ |\psi^{(0)}_- )\right] \hat Q^{(0)}_{rel} \label{q1}.
\end{eqnarray}
Before  calculating the coefficients of the differential equation (\ref{q1}) we remark that their scalings with $N$ are different. Indeed one of the terms in front of $\hat P^{(0)}_{rel}$ is proportionnal to the quantity of the type $(\phi^{(0)}_{+} |\phi^{(0)}_{-})$ which is divergent in the hydrodynamic limit, and in order to calculate it, corrections to the hydrodynamic limit have to be taken into account. The correction of the divergence of the integral is achieved by introducing the healing lenght $\xi$ of the condensate defined as, $\xi = {\left(\frac{2m^2 \omega^2 r_0}{\hbar^2}\right)}^{-\frac{1}{3}}$ (see \cite{Stringari}).
At the first order in the difference of the traps the scaling of this term with $N$ is like $v \ln N$ although the others scale as $v$. Since we are working in the hydrodynamic approximation which is valid for the great values of $N$ (see \cite{StringariTF}), we can  neglect the contribution of the terms proportionnal to $\hat Q^{(0)}_{rel}$ in comparison with those  proportionnal to $\hat P^{(0)}_{rel}$ and in fact forget all the terms which do not have the form of a ``$\phi \phi$'' product. The final  result is the following:
\begin{eqnarray}
\hat Q^{(1)}_{rel}(t) &\simeq& \frac{6}{5} v \ln(\frac{2\xi}{r_0})\frac{N }{\sqrt{N_+ N_-}}{\rm sin}(2\lambda t +\Delta\theta_0)\hbar\hat P^{(0)}_{rel}
\end{eqnarray}
where $\Delta \theta_0 = (B^{(0)}_+ -B^{(0)}_-)/\hbar$ is the difference between the phase of the zero order wave functions (see (\ref{zero+},\ref{zero-})). In the first order in $v$ the fluctuations of the relative phase are an oscillating function of frequency $2 \lambda$, provided $r_0$ is constant in time (the latter assumption means that at the time $t=0$ the densities of the A and B atoms have the stationary form, and the dynamics is triggered only by the phase difference). 
It can be noticed that $v \propto r_0$ and that in the hydrodynamic limit we have $r_0 \propto N^{2/5}$ so that $v \propto N^{2/5}$ too. It means that our pertubative developement in the parameter $v$ is valid only for values of $\delta V,\lambda$ and $N$ fullfilling the condition $v\ll 1$. 
At the first order in $v$ the first order correlation function is then a $\frac{\pi}{\lambda}$ periodic function. The physical interpretation of this result is that the relative phase  between the two condensates exhibits some fluctuations periodic in time. This dynamical behavior could be compared to that of a squeezed state of an harmonic oscillator undergoing free evolution.

In order to find a diffusion term we have thus to go to the second order in the development.
We find (retaining only the coefficients presenting a scaling with $N$ as $v^2 \ln N$, where the logarithmic dependance with $N$ can only come from integrals involving a product between two functions $\phi$) :
\begin{eqnarray}
\frac{d \hat Q^{(2)}_{rel}}{dt} &=& -16\left[{\rm Re}(\phi^{(0)}_+|\phi^{(0)}_-)\right] \hat P^{(1)}_{rel} -16\left[{\rm Re}(\phi^{(1)}_+|\phi^{(0)}_-) +{\rm Re}(\phi^{(0)}_+|\phi^{(1)}_-)\right] \hat P^{(0)}_{rel}
+\frac{2}{\hbar}\left[ {\rm Im}(\psi^{(0)}_+ |\phi^{(0)}_- ) - {\rm Im}(\phi^{(0)}_+ |\psi^{(0)}_- )\right] \hat Q^{(1)}_{rel} 
\end{eqnarray}
The evaluation of the first order correction of the $\phi$ functions is made in the Appendix. For the final result we have only retained the part of the operators proportionnal to $\hat P^{(0)}_{rel}$ (assuming that the initial fluctuations with the relative phase are small compared to those of the relative number of atoms) and the part giving effectively the diffusion term. We then have:
\begin{eqnarray}
\hat Q^{(2)}_{rel}(t) \simeq \frac{12}{25} v^2 \ln (\frac{2\xi}{r_0})\frac{N\Delta N}{N_+N_-}\hbar \lambda \hat{P^{(0)}_{rel} }t+...\label{diff1}
\end{eqnarray}
where $\Delta N =N_+ -N_-$.
This result shows that at the second order $\hat Q_{rel}$ is growing linearly in time. The first order correlation function will then present the usual Gaussian decay in time :
\begin{eqnarray}
<\hat\Psi^{\dag}_+(t)\hat\Psi_-(t)> \propto {\rm exp}\left(-\frac{t^2}{\tau^2_D}\right).
\end{eqnarray}
with 
\begin{equation}
\tau^{-1}_D = -\frac{12}{25} v^2 \ln (\frac{2\xi}{r_0})\frac{N^2}{N_+N_-} \lambda \sqrt{<\hat P^{(0)2}_{rel}>}
\end{equation}
assuming that the mean value of the relative number of atoms is fixed which means that $<\hat P^{(0)}_{rel}>=0$.

\section{Conclusions}
Generelazing the method previously developed in Ref.\cite{revue}, we have made an analysis of the dynamics of two coupled condensates adopting a complete field theoretical approach.
We have derived the evolution equations of the operators associated with the fluctuations of the total and relative number of atoms and phase of the condensates. The first result is that similarly to the static case (see \cite{revue}), the dynamics of the relative operators is decoupled to that of the total operators.
Using these evolution equations it is possible to express the  equal time first order correlation function and show that its behavior is drastically influenced by the difference between the two trap potentials. In the case of harmonic traps with equal frequencies the result is that there is no dephasing between the two phases of the two condensates. The relative phase fluctuations remain constant in time. But if the traps are different the phases present a dephasing with time. The dephasing process is due to two effects: the first is the non-linearity of the dynamics of the system, the second is the  usual spreading of a wave packet in quantum mechanics during its propagation.
We have shown that when the relative number of atoms is allowed to fluctuate around a fixed mean value, these fluctuations  induce a diffusion of the relative phase. We have analysed the dephasing time for the case of two harmonic traps having different frequencies in the limit where the ratio $v = \frac{m\delta \omega^2 r_0}{4\hbar \lambda}$ is small compared to 1 using a perturbative approach, and shown that this dephasing time is of the second order in $v$. It can be expressed as:
\begin{equation}
\tau^{-1}_D = -\frac{12}{25} v^2 \ln (\frac{2\xi}{r_0})\frac{N \Delta N}{N_+N_-} \lambda \sqrt{<\hat P^{(0)2}_{rel}>}.
\end{equation}
                                                                                                                                 
It is worth noticing that the characteristic dephasing time depends on the healing lenght of the condensates, which indicates that the phenomenon takes its source in surface effects.

To conclude we remark that the method developed here could as well be applied to the case of two harmonic traps shifted with equal frequency (see \cite{jimmy}) and could be generalized to the case of the mixing of three condensates (see for instance \cite{bigelow} and references therein). Into the latter case, the method allows to calculate the spin mixing time of different spinor components of the Bose-Einstein condensate.

We thank Y. Castin and R. Dum for enlightening discussions.

\subsection{Appendix}
In order to evaluate coefficients in the differential evolution equation of the operator $\hat Q_{rel}$ we have to calculate the first order correction to the $\psi_{\pm}$ functions and of the $\phi_{\pm}$ functions.
To this aim we decompose the functions as:
\begin{eqnarray}
\psi_+ (\vec r ,t) &=& \psi^{(0)}_+ (r,t)+ A_+ (\vec r\,)e^{2i\lambda t} + B_+ (\vec r\,)e^{-2i\lambda t},          \\
\psi_- (\vec r ,t) &=& \psi^{(0)}_- (r,t) +A_- (\vec r\,)e^{2i\lambda t} + B_- (\vec r\,)e^{-2i\lambda t},          \\
\phi_+ (\vec r ,t) &=& \phi^{(0)}_+ (r,t) +C_+ (\vec r\,)e^{2i\lambda t} + D_+ (\vec r\,)e^{-2i\lambda t},          \\
\phi_- (\vec r ,t) &=& \phi^{(0)}_- (r,t) +C_- (\vec r\,)e^{2i\lambda t} + D_- (\vec r\,)e^{-2i\lambda t}.
\end{eqnarray}
In this way we take into account only the most relevant Fourier's components of the function.
We then introduce these forms into the Eqs.(\ref{NNLSE1},\ref{NNLSE2}) and linearize them in terms of the corrections to the zero order functions.
As the inhomogeneous terms in the equations for the complex functions $A_{\pm},B_{\pm},C_{\pm},D_{\pm}$ are proportionnal to $\delta V$ these functions are the first order corrections that we are searching.
For the first order corrections of the $\psi$ function we have the following system of time independant differential equations:
\begin{eqnarray}
{\rm M} 
\left[ \matrix{A_+\cr A^*_+ \cr B_+ \cr B^*_+ \cr A_- \cr A^*_- \cr B_- \cr B^*_- \cr }\right]= 
-\delta V\left[ \matrix{ 0\cr 0\cr  \psi^{(0)}_- \cr \psi^{(0)*}_- \cr \psi^{(0)}_+ \cr \psi^{(0)*}_+ \cr 0 \cr 0 \cr} \right] \label{correcpsi} 
\end{eqnarray}
where M is the following 8 by 8 matrix:
\begin{equation}
{\rm M} = \left[\matrix{ {\rm M_{++}}& {\rm M_{+-}}\cr {\rm M^{\dag}_{+-}}& {\rm M_{--}}\cr}\right]
\end{equation}
with:
\begin{eqnarray}
{\rm M_{++}}&=& \left[\matrix{
{\cal L}^+_0 +2\hbar \lambda& 0&0&Nu_0 \psi^{(0)2}_+\cr
0&{\cal L}^+_0 +2\hbar \lambda&Nu_0 \psi^{(0)*2}_+&0\cr
0 & N u_0 \psi^{(0)2}_+ & {\cal L}^+_0 -2\hbar \lambda & 0 \cr
Nu_0 \psi^{(0)*2}_+ & 0 & 0 & {\cal L}^+_0 -2\hbar \lambda \cr}\right], \\
\\
{\rm M_{--}}&=& \left[ \matrix{{\cal L}^-_0 +2\hbar \lambda & 0 & 0 & N u_0 \psi^{(0)2}_- \cr
 0 & {\cal L}^-_0 +2\hbar \lambda  & Nu_0 \psi^{(0)*2}_- & 0 \cr
0 &  N u_0 \psi^{(0)2}_- & {\cal L}^-_0 -2\hbar \lambda & 0 \cr
 Nu_0 \psi^{(0)*2}_- & 0 & 0 &  {\cal L}^-_0 -2\hbar \lambda \cr}\right],\\
\\
{\rm M_{+-}}&=&\left[\matrix{Nu_0 \psi^{(0)*}_- \psi^{(0)}_+&0&0&Nu_0 \psi^{(0)}_- \psi^{(0)}_+ \cr
Nu_0 \psi^{(0)}_- \psi^{(0)*}_+&Nu_0 \psi^{(0)*}_- \psi^{(0)*}_+&0&0\cr
0 &  Nu_0 \psi^{(0)}_- \psi^{(0)}_+ &  Nu_0 \psi^{(0)*}_- \psi^{(0)}_+ & 0 \cr
Nu_0 \psi^{(0)*}_- \psi^{(0)*}_+ & 0 & 0 &  Nu_0 \psi^{(0)}_- \psi^{(0)*}_+ \cr}\right],
\end{eqnarray}
where  we denoted by ${\cal L}^{+,-}_0$ the operator ${\cal L}_0 + u_0 \rho^{(0)}_{+,-}$ with ${\cal L}_0 = {\cal L} +u_o \rho^{(0)}$.
In fact as we are working in the hydrodynamic approximation we can neglect the result of the action of the Laplacian in ${\cal L}$ on the first order corrections. That means that the above differential system becomes in fact an algebraic system of equations.
We can write the matrix M in the following form:
\begin{equation}
{\rm M} = \Lambda + {\rm M_0},
\end{equation}
where $\Lambda$ is the following diagonal matrix:
$\Lambda = \left[ \matrix{J & 0 \cr 0 & J} \right]$
with
\begin{equation} 
J = \left[ \matrix{ 2\hbar \lambda & 0 & 0 & 0\cr
0&2\hbar \lambda & 0 & 0 \cr
0& 0 & -2\hbar \lambda & 0 \cr
0& 0 & 0 & -2\hbar \lambda \cr }\right].
\end{equation}
Proceeding in this way, the formal solution of the system of equations (\ref{correcpsi}) is:
\begin{equation}
\left[\Psi^{(1)}\right]= -\frac{1}{\Lambda + {\rm M_0}}\-\delta V \left[\Psi^{(0)}\right],
\end{equation}
where we use a vector notation.
In the vinicity of $r_0$ all the coefficients of the matrix ${\rm M_0}$ go to zero, so that the first order correctiontof the $\psi $ function have asymptotically the same behavior as the zero order parts.
Namely, we can write for $r\rightarrow r_0$:
\begin{equation}
\left[\Psi^{(1)}\right]\simeq -\frac{1}{\Lambda}\delta V \left[\Psi^{(0)}\right] \label{psi1}
\end{equation}
Using the same approach we find the following system for the first order correction to the $\phi$ functions:
\begin{eqnarray}
{\rm M}\left[\matrix{ C_+\cr C^*_+ \cr D_+ \cr D^*_+ \cr C_- \cr C^*_- \cr D_- \cr D^*_- \cr }\right] =
\left[ \matrix{ T_1 \cr T^*_1 \cr T_2  \cr T^*_2 \cr\hat \Pi_{+-}T_1  \cr \hat \Pi_{+-}T^*_1 \cr  \hat \Pi_{+-}T_2 \cr\hat \Pi_{ +-}T^*_2 \cr } \right]-\delta V\left[ \matrix{ 0\cr 0 \cr \phi^{(0)}_- \cr\phi^{(0)*}_- \cr \phi^{(0)}_+ \cr\phi^{(0)*}_+ \cr 0 \cr 0 \cr} \right],\label{correc}
\end{eqnarray}
where $\hat \Pi_{+-}$ is the operator interchanging the signs $+$ and $-$, and the terms $T_1$ and $T_2$ are:
\begin{eqnarray}
T_1&=&  \left[N\tilde u^{(0)} A_+ -2Nu_0 A_+(\phi^{(0)}_+ \psi^{(0)*}_+ + \phi^{(0)*}_+ \psi^{(0)}_+)-Nu_0 A_+ (\phi^{(0)}_- \psi^{(0)*}_- + \phi^{(0)*}_- \psi^{(0)}_-)\right]-Nu_0 A_- (\phi^{(0)}_+ \psi^{(0)*}_+ + \phi^{(0)*}_+ \psi^{(0)}_+)\nonumber \\
&-& 2Nu_0 B^*_+ \psi^{(0)}_+ \phi^{(0)}_+ +  Nu_0 B^*_- (\psi^{(0)}_+ \phi^{(0)}_- +\psi^{(0)}_- \phi^{(0)}_+) \\
T_2 &=& \left[N\tilde u^{(0)} B_+ -2Nu_0 B_+(\phi^{(0)}_+ \psi^{(0)*}_+ + \phi^{(0)*}_+ \psi^{(0)}_+)-Nu_0 B_+ (\phi^{(0)}_- \psi^{(0)*}_- + \phi^{(0)*}_- \psi^{(0)}_-)\right]-Nu_0 B_- (\phi^{(0)}_+ \psi^{(0)*}_+ + \phi^{(0)*}_+ \psi^{(0)}_+)\nonumber \\
&-& 2Nu_0 A^*_+ \psi^{(0)}_+ \phi^{(0)}_+ +  Nu_0 A^*_- (\psi^{(0)}_+ \phi^{(0)}_- +\psi^{(0)}_- \phi^{(0)}_+)
\end{eqnarray}

We are mainly interested by the first order correction to the $\phi$ functions which will make lead to a divergence in the integrals of the form ${\rm Re}(\phi^{(1)}_{\pm}|\phi^{(0)}_{\mp})$ in the hydrodynamic limit.
 The formal solution for the first order correction of the system (\ref{correc}) is:
\begin{equation}
\left[\Phi^{(1)}\right]= \frac{1}{\Lambda + {\rm M_0}}\left( \left[T\right]-\delta V \left[\Phi^{(0)}\right] \right),
\end{equation}
where the notations are the same as above.
The first vector involves only products of the form $\psi^{(1)}_{\pm}\psi^{(0)}_{\pm}\phi^{(0)}_{\pm}$, so that in the integrand it will contribute in the form $\psi^{(1)}_{\pm}\psi^{(0)}_{\pm}\phi^{(0)}_{\pm}\phi^{(0)}_{\pm}$ whose behavior in the vinicity of the boundary is continuous, since the first order corrections to the $\psi$ functions behave as the zero ordercounterpart (see (\ref{psi1})). 
For $r\rightarrow r_0$ we have: 
\begin{equation}
\left[\Phi^{(1)}\right]\simeq -\frac{1}{\Lambda}\delta V \left[\Phi^{(0)}\right], \label{sol}
\end{equation}
which shows that the first order corrections of the $\phi$ function have the same kind of divergence than the zero order parts at the boundary, and that the integrals of the kind ${\rm Re}(\phi^{(1,0)}_{\pm}|\phi^{(0,1)}_{\pm})$ will scale as $v^2\ln N$. In the large $N$ limit, only these logarithmic divergent terms must effectively be used in the evolution equation of $\hat Q^{(2)}_{rel}$.
Moreover these most relevant parts of the first order correction of the $\phi$ functions can be calculated from Eqn.(\ref{sol}). 
It then appears that the integrals ${\rm Re}(\phi^{(1)}_{+}|\phi^{(0)}_{-})$ and ${\rm Re}(\phi^{(0)}_{+}|\phi^{(1)}_{-})$ are constant in time, and take part in the diffusion process. For instance we have:
\begin{equation}
{\rm Re}(\phi^{(1)}_{+}|\phi^{(0)}_{-})= -\frac{3}{40}v^2\ln(\frac{\xi}{r_0})\hbar \lambda \frac{N}{N_-}
\end{equation}
Careful calculation of these terms allows to derive Eqs.(\ref{diff1}).

\end{document}